\documentclass[sigconf]{acmart}

\usepackage{listings}
\usepackage{framed}
\usepackage{tabularx}
\usepackage{float}
\usepackage{fontawesome}

\usepackage{booktabs}
\usepackage{longtable}
\usepackage{xspace}

\definecolor{shadecolor}{HTML}{ffffe6}
\usepackage{graphicx}
\usepackage{textcomp}
\usepackage{algorithm}
\usepackage{algpseudocode}
\usepackage{xcolor}
\usepackage{url}
\usepackage{enumitem}
\usepackage{comment}
\usepackage{array}

\usepackage{multirow}
\usepackage{makecell}
\usepackage{afterpage}
\usepackage{float}
\usepackage{graphicx}
\usepackage{tabularx}
\usepackage{hyperref}
\usepackage{longtable}
\usepackage{multirow}
\usepackage{array}
\usepackage[justification=centering]{caption}
\usepackage{tikz}
\usetikzlibrary{positioning}
\usepackage{soul}
\usepackage{hyperref}
\usepackage{tablefootnote}

\definecolor{source}{gray}{0.95}
\definecolor{highlight}{gray}{0.9}
\definecolor{bblue}{HTML}{4F81BD}
\definecolor{rred}{HTML}{C0504D}
\definecolor{ggreen}{HTML}{9BBB59}
\definecolor{ppurple}{HTML}{9F4C7C}

\usepackage{textcase}  
\usepackage{multirow}

\definecolor{source}{gray}{0.9}

\copyrightyear{2024}
\acmYear{2024}
\setcopyright{rightsretained}
\acmConference[Preprint Version, ESEM '24]{Proceedings of the 18th ACM / IEEE International Symposium on Empirical Software Engineering and Measurement}{October 24--25, 2024}{Barcelona, Spain}
\acmBooktitle{Proceedings of the 18th ACM / IEEE International Symposium on Empirical Software Engineering and Measurement (ESEM '24), October 24--25, 2024, Barcelona, Spain}\acmDOI{10.1145/3674805.3695408}
\acmISBN{979-8-4007-1047-6/24/10}

\setcopyright{none}

\begin{document}

\title{ChatGPT's Potential in Cryptography Misuse Detection: A Comparative Analysis with Static Analysis Tools}

\author{Ehsan Firouzi}
\affiliation{%
  \institution{Technische Universität Clausthal}
  \country{Germany}
}

\author{Mohammad Ghafari}
\affiliation{%
  \institution{Technische Universität Clausthal}
  \country{Germany}
}

\author{Mike Ebrahimi}
\affiliation{%
  \institution{CUBE Global}
  \country{Australia}
}

\begin{abstract}

The correct adoption of cryptography APIs is challenging for mainstream developers, often resulting in widespread API misuse. Meanwhile, cryptography misuse detectors have demonstrated inconsistent performance and remain largely inaccessible to most developers.
We investigated the extent to which ChatGPT can detect cryptography misuses and compared its performance with that of the state-of-the-art static analysis tools.
Our investigation, mainly based on the CryptoAPI-Bench benchmark, demonstrated that ChatGPT is effective in identifying cryptography API misuses, and with the use of prompt engineering, it can even outperform leading static cryptography misuse detectors.

\end{abstract}



\maketitle

\section{Introduction}
Cryptography plays a vital role in software security by safeguarding data confidentiality. However, improper adoption (i.e., misuses) is prevalent~\cite{hazhirpasand2021_a, hazhirpasand2020, 9609232, firouzi2024time}, and these misuses occur regardless of developer experience~\cite{8870184}.
In response to these challenges,  researchers have invested significant effort in building misuse detection tools \cite{CogniCrypt, CryptoGuard,piccolboni2021} and designing more usable APIs~\cite{Firouzi2024SafEncrypt, Kafader2021}.

The recent surge in the use of Large Language Models (LLMs), such as ChatGPT, to facilitate software development has motivated researchers to explore these models for enhancing security as well~\cite{Kavian2024-lv, Catherine24}.
In this paper, we delve into this exciting prospect by investigating the capabilities of ChatGPT in detecting security risks specifically related to the Java Cryptography Architecture (JCA).
Therefore, we asked the following research question (RQ): 
\emph{How well does ChatGPT perform in detecting Java cryptography misuse?}

We conducted a rigorous evaluation primarily based on CryptoAPI-Bench, a benchmark specifically designed for Java cryptography analysis~\cite{8901573}.
We assessed ChatGPT's capability in detecting a variety of cryptography misuses and compared its performance to that of the state-of-the-art cryptography misuse detectors.\footnote{We used GPT-3.5-Turbo, the default model for ChatGPT at the time this work was conducted.}

The initial results revealed that ChatGPT achieved an average F-measure of 86\% across 12 misuse categories. When compared to CryptoGuard~\cite{CryptoGuard}, the leading misuse detector~\cite{AfroseDetectionTools, AutoDetecJavaCrypto}, ChatGPT outperformed in five categories (notably, 92.43\% versus 76.92\% for detecting predictable keys), while CryptoGuard was superior in four categories (notably, 76.92\% versus 59.14\% for detecting low iteration counts). However, after applying prompt engineering techniques, we were able to enhance ChatGPT's detection capabilities, resulting in an improved average F-measure of 94.6\%. This adjustment enabled ChatGPT to outperform the state-of-the-art crypto misuse detector, with superior performance in 10 categories and nearly identical results in the remaining two.
Finally, we confirmed the generalizability of our engineered prompts by evaluating them using a different benchmark, called CAMBench~\cite{schlichtig2022cambench}.

In summary, our investigations highlight ChatGPT's promising potential for detecting cryptography misuse.

The remainder of this paper is organized as follows.
In Section~\ref{sec:background}, we provide an overview of existing cryptography misuse detectors and relevant cryptography benchmarks.
In Section~\ref{sec:methodology}, we describe our methodology.
In Section~\ref{sec:result}, we present the results and discuss them in Section~\ref{sec:Discussion}. 
We discuss the threats to validity in Section~\ref{sec:ThreatstoValidity} and conclude the paper in Section~\ref{sec:Conclusion}.   


\begin{table*}[t]
\caption{CryptoAPI Benchmark test coverage for cryptography misuses}
\scalebox{0.92}{
\begin{tabular}{p{3.8cm}p{4.9cm}p{1.7cm}p{1.9cm}p{2.5cm}l}
\hline
Misuses                               & Description                                                                                                               & CWE\_ID           & Attack                   & JCA APIs                                        & \#Test Cases \\ \hline
\begin{tabular}[c]{@{}l@{}}Weak   symmetric Encryption\\ Algorithm\end{tabular} & \begin{tabular}[c]{@{}l@{}}Vulnerable algorithms e.g., DES,  \\  Blowfish,    RC4, RC2, IDEA\end{tabular}           & CWE-327          & Brute-force  &        Cipher                                          & 36          \\
Weak Encryption Mode                  & Using ECB                                                                                                            & CWE-327          & CPA    & Cipher                                          & 8           \\
Predictable Key                       & Constant or Hard-coded                                                                                               & CWE-798          & Predictability           & SecretKeyspec                                   & 10          \\
Static Salt                           & Using   Static Salt for key derivation                                                                               & CWE-330          & CPA & PBEKeySpec   
 PBEParameterSpec                   & 9           \\
Low Iteration Count (PBE)              & Lower than 1000                                                                                                      & CWE-330, CWE-326  & Brute-force              & PBEKeySpec PBEParameterSpec                     & 9           \\
Weak Random generation                & Using weak Random function for   generating secret key, Predictable seed                                             & CWE-330          & Predictability           & SecureRandom                                    & 19          \\
Perdictable Password                  & Using Static (= Constant)   Password                                                                       & CWE-259          & Predictability           & PBEKeySpec                                      & 11          \\
Static IV                             & Hardcoded or Constant IV                                                                                             & CWE-330, CWE-798 & CPA & IVparameterSpec                                  & 10          \\
Weak Password in keyStore             & Hardcoded or Constant Password                                                                                     & CWE-798          & Predictability           & KeyStore                                        & 10          \\
Vulnerable SSL/TSL                    & Verifing host names or   certificates  in SSL in trivial ways or   manually change the hostname verifier, Using http & CWE-295, CWE-757 & MITM                     & X509TrustManager SSLSocket    
     HostnameVerifier & 14          \\
Insecure   Asymmetric Ciphers         & RSA with keysize<1024                                                                                                & CWE-326          & Brute-force              & Cipher                                          & 6           \\
Weak Hash Function                    & Using broken hash functions   (SHA1, MD2, MD5, ...)                                                                   & CWE-328          & Brute-force              & MessageDigest                                   & 29          \\ \hline
Total                                 &                                                                                                                      &                  &                          &                                                 & 171         \\ \hline
\end{tabular}
}
\label{tbl:CryptoAPIMisuses}
\end{table*}
\section{Java Cryptography}
\label{sec:background}

We present existing tools to uncover cryptography misuses as well as the established benchmarks to compare such tools.   

\subsection{Misuse Detectors}
Java Cryptography Architecture (JCA) is a Java framework for cryptographic operations like encryption, decryption, signing, and verification. While JCA offers comprehensive tools, correct usage is vital to ensure security, as improper implementation can lead to vulnerabilities.
Misuse of JCA is common, as evidenced by various studies~\cite{hazhirpasand2020, nadi2016}. Incorrect usage can compromise security through weak keys, inadequate algorithm selection, improper handling of salts and nonces, or incorrect signature validation. To mitigate these risks, static analysis tools exist to identify cryptography misuse, e.g., SpotBugs with FindSecBugs plugin \cite{find_secbugs_2024}, CogniCrypt \cite{b13}, CryLogger \cite{piccolboni2020}, and CryptoGuard \cite{rahaman2019cryptoguard}.

\subsection{Benchmarks}
Evaluating these tools requires robust benchmarks with realistic misuse patterns. 
In the realm of JCA misuse detection tools, several benchmarks exist (e.g., OWASP Benchmark \cite{owasp_benchmark_2024}, MUBench \cite{ANNNM16}, CryptoAPI-Bench \cite{8901573}, CAMBench \cite{schlichtig2022cambench}), each with its limitations. We chose CryptoAPI-Bench as it is specifically designed for cryptographic analysis. In contrast, other benchmarks often include cryptography as just a subset of broader security concerns. Additionally, the performance results of various static analysis tools on CryptoAPI-Bench are well-documented in academic literature. This allows us to compare ChatGPT's results with those of existing tools, providing a comprehensive analysis of its capabilities. 
For checking the generalizability of our prompt optimization, we used CAMBench—a new, comprehensive benchmark for cryptographic API misuse detection tools. However, we did not use it as the base due to the lack of evaluation results for existing tools.

\textbf{CryptoAPI-Bench}.
This benchmark comprises 171 test cases utilizing the JCA and Java Secure Socket Extension (JSSE) APIs. Among these, 136 intentionally contain cryptographic API misuses, while the remaining 35 use the APIs correctly.

The test cases address various scenarios, each with a distinct purpose:

\emph{1. Basic Cases:} Focus on vulnerabilities within a single method, providing a foundational assessment of a tool's ability to detect straightforward cryptographic flaws.

\emph{2. Advanced Cases:} This category evaluates a tool's ability to identify vulnerabilities from various sources like methods, classes, variables, or conditional statements.

\emph{2.1. Interprocedural:} Involve vulnerabilities spanning multiple methods, challenging the tool to trace complex method invocations.

\emph{2.2. Field-sensitive:} Examine vulnerabilities dependent on data flow across different fields of the same object, evaluating the tool's precision in analyzing object-oriented structures.

\emph{2.3. Combined:} Present sophisticated scenarios merging interprocedural and field-sensitive aspects, requiring holistic vulnerability detection.

\emph{2.4. Path-sensitive:} Involve vulnerabilities based on conditional branches, testing the tool's capability to navigate diverse program paths.

\emph{2.5. Miscellaneous:} Include cases with irrelevant constraints and interfaces, testing the tool's robustness against diverse inputs and scenarios beyond typical use cases.

\emph{2.6. Multiple Class:} Involve vulnerabilities originating from external Java classes, simulating real-world situations where cryptographic weaknesses may spread across different program modules.

The misuses covered by cryptoAPI-bench are listed in Table \ref{tbl:CryptoAPIMisuses}. It also includes the description of each misuse, corresponding  CWE-IDs, potential attacks, relevant APIs, and the number of test cases associated with each misuse.

Previous studies~\cite{AfroseDetectionTools, AutoDetecJavaCrypto} evaluated several tools using CryptoAPI-Bench and found CryptoGuard to be the most effective one, making this tool our baseline for comparison.

Table~\ref{tbl:compare} presents the precision, recall, and F-measure metrics for the CryptoGuard tool.
These results are extracted from the full benchmark results reported in~\cite{CryptoGuard}. However, what the authors showed in the original paper covers only a subset of the benchmark (6 rules) rather than all rules.

\begin{table*}
\caption{The performance of CryptoGuard versus ChatGPT}
\scalebox{0.95}{

\begin{tabular}{|l|lll|lll|lll|}
\hline
\multirow{2}{*}{Misuse Category} & \multicolumn{3}{c|}{CryptoGuard}                                                            & \multicolumn{3}{c|}{ChatGPT}                                                                & \multicolumn{3}{c|}{ChatGPT with engineered prompts}                                          \\ \cline{2-10} 
                                                       & \multicolumn{1}{c}{Percision} & \multicolumn{1}{c}{Recall} & \multicolumn{1}{c|}{F-Measure} & \multicolumn{1}{c}{Percision} & \multicolumn{1}{c}{Recall} & \multicolumn{1}{c|}{F-Measure} & \multicolumn{1}{c}{Percision} & \multicolumn{1}{c}{Recall} & \multicolumn{1}{c|}{F-Measure} \\ \hline
\multicolumn{1}{|l|}{Weak Encryption Algorithm}        & 85.71\%                         & 100.00\%                     & 92.31\%                          & 78.00\%                         & 83.00\%                      & 80.42\%                          & 96.00\%                         & 100.00\%                     & 97.96\%                          \\
\multicolumn{1}{|l|}{Weak Encryption Mode}             & 85.71\%                         & 100.00\%                     & 92.31\%                          & 85.00\%                         & 100.00\%                     & 91.89\%                          & 95.00 \%                        & 100.00 \%                    & 97.44\%                          \\
\multicolumn{1}{|l|}{Predictable Key}                  & 83.33\%                         & 71.43\%                      & 76.92\%                          & 90.00\%                         & 95.00\%                      & 92.43\%                          & 95.00\%                         & 95.00\%                      & 95.00\%                          \\
\multicolumn{1}{|l|}{Static Salt}                      & 85.71\%                         & 85.71\%                      & 85.71\%                          & 83.00\%                         & 95.00\%                      & 88.60\%                          & 90.00\%                         & 95.00\%                      & 92.43\%                          \\
\multicolumn{1}{|l|}{Low Iteration Count (PBE)}         & 83.33\%                         & 71.43\%                      & 76.92\%                          & 48.00\%                         & 77.00\%                      & 59.14\%                          & 81.00\%                         & 90.00\%                      & 85.26\%                          \\
\multicolumn{1}{|l|}{Weak Random generation}           & 85.71\%                         & 80.00\%                      & 82.76\%                          & 83.00\%                         & 97.00\%                      & 89.46\%                          & 90.00\%                         & 100.00\%                     & 94.74\%                          \\
\multicolumn{1}{|l|}{Perdictable password}             & 87.50\%                         & 87.50\%                      & 87.50\%                          & 78.00\%                         & 87.00\%                      & 82.25\%                          & 82.00\%                         & 91.00\%                      & 86.27\%                          \\
\multicolumn{1}{|l|}{Static IV}                        & 87.50\%                         & 87.50\%                      & 87.50\%                          & 88.00\%                         & 100.00\%                     & 93.62\%                          & 92.00\%                         & 100.00\%                     & 95.83\%                          \\
\multicolumn{1}{|l|}{Predictable Password in KeyStore} & 88.00\%                         & 100.00\%                     & 93.62\%                          & 87.50\%                         & 100.00\%                     & 93.33\%                          & 95.00\%                         & 100.00\%                     & 97.44\%                          \\
\multicolumn{1}{|l|}{Vulnerable SSL/TSL}               & 92.00\%                         & 100.00\%                     & 95.83\%                          & 87.00\%                         & 78.00\%                      & 82.25\%                          & 92.00\%                         & 100.00\%                     & 95.83\%                          \\
\multicolumn{1}{|l|}{Insecure Asymmetric Ciphers}      & 80.00\%                         & 80.00\%                      & 80.00\%                          & 83.00\%                         & 100.00\%                     & 90.71\%                          & 93.75\%                         & 100.00\%                     & 96.77\%                          \\
\multicolumn{1}{|l|}{Weak Hash Function}               & 86.00\%                         & 100.00\%                    & 92.47\%                          & 86.74\%                         & 100.00\%\                     & 92.90\%                          & 96.00\%                         & 100.00\%                     & 97.96\%                          \\ \hline
\end{tabular}
}
\label{tbl:compare}
\end{table*}

\section{Methodology}
\label{sec:methodology}

We assessed ChatGPT’s proficiency in detecting JCA misuse, and then improved its detection capabilities through prompt engineering. Lastly, we conducted a re-evaluation to confirm the results of our prompt optimization.

\subsection{ChatGPT's JCA Misuse Detection}
The objective of this stage was to evaluate ChatGPT's ability to identify misuse within the JCA using the CryptoAPI-Bench benchmark dataset. To ensure a comprehensive analysis, we followed four steps:

Step 1. Research and Compilation of Security Violation Rules: We conducted an extensive review of the latest advancements in crypto misuse analysis. This review included consulting tools such as CryLogger~\cite{piccolboni2020}  and CogniCrypt~\cite{b13}  and examining insights from recent scholarly article \cite{firouzi2024time}. From this research, we compiled a detailed list of security violation rules that serve as guidelines for identifying potential vulnerabilities in cryptographic implementations.

Step 2. Analysis of CryptoAPI-Bench Gaps: We examined CryptoAPI-Bench to identify scenarios that lacked test cases, based on the security rules identified in Step 1. For each identified gap, we sourced three relevant code snippets from Stack Overflow that demonstrated the insecure patterns derived from our compiled violation rules.

Step 3. Security Evaluation of Test Cases: We prompted ChatGPT (GPT-3.5-turbo) to evaluate the security of each CryptoAPI-Bench test case using the query: ``Please assess the security of the provided code and identify any vulnerabilities''. Recognizing the potential variability in responses from ChatGPT for identical prompts at different times, we issued this query three times to ensure reliability in the results.

Step 4. Assessment of Extracted Code Snippets: We repeated the security evaluation process for the code snippets sourced in Step 2 to further assess ChatGPT’s detection capabilities.

In the end, we identified specific test cases and code snippets where ChatGPT exhibited limitations in detecting misuse effectively.

\subsection{Prompt Engineering}

To enhance ChatGPT's detection abilities, we adopted a systematic approach detailed in the following steps:

Step 1. We searched Google Scholar for recent research papers that contained keywords such as  ``ChatGPT'' or ``Prompt Engineering''.
We limited our investigation to papers published over the last three years (2022-2024) in top venues (core ranking A or A+) or those on arXiv with more than 5 citations.
Additionally, we manually checked the conference programs for top-tier venues such as
ICSE, FSE, ASE and ESEM to find related papers. 
From these sources, we collected prompt patterns acknowledged to enhance the quality of ChatGPT's responses.

Step 2. Development of Optimized Prompts: Building on insights from the recent scholarly article \cite{firouzi2024time} and the security violation rules identified in the previous stage, we utilized the prompt patterns established in Step 1. This led to the creation of optimized prompts specifically tailored for each misuse category.

\subsection{Evaluation} 
We evaluated the optimized prompts using the CryptoAPI-Bench benchmark to compare the results with ChatGPT's performance without prompts. Additionally, to ensure generalizability, we also conducted an additional evaluation using a different benchmark.
Specifically, for each test case where ChatGPT initially performed weakly, we executed new test cases from CAMBench to ensure the robustness of our findings and to mitigate any bias.

\section{Result}
\label{sec:result}
We investigated the performance of ChatGPT in uncovering cryptography misuses in JCA and applied prompt engineering to enhance its effectiveness. Table~\ref{tbl:compare} shows the performance results of CryptoGuard, ChatGPT, and ChatGPT with engineered prompts based on misuse categories, while Table \ref{tab:BasicvsAdvanced} compares ChatGPT's performance with and without prompt engineering in both basic and advanced test cases.

\subsection{ChatGPT's JCA Misuse Detection}
\label{subsec:RQ1}

\subsubsection{Weak symmetric Encryption Algorithm:}
While the ChatGPT evaluation result suggests that it does not match up to CryptoGuard's performance in weak algorithm detection, it still demonstrates commendable proficiency, particularly in detecting basic cases. Additionally, it exhibits a robust ability to identify DES even in complex scenarios like interprocedural and field-sensitive contexts. However, our analysis reveals a notable issue where it mistakenly did not categorize IDEA \footnote{https://chatgpt.com/share/6882a99d-15b3-4523-9d7b-fc3f6647971e} and Blowfish\footnote{https://chat.openai.com/share/61707de2-8e59-4b2e-aa16-7cd39b8a240f
} as weak algorithms. 

\subsubsection{Weak Encryption Mode:}
Remarkably, both ChatGPT and CryptoGuard exhibited the same precision and recall in detecting the insecure Electronic Codebook (ECB) encryption mode.  In a specific test case where no security risks were present, ChatGPT not only confirmed the absence of insecurity but also issued a warning: ``\emph{Although the class name suggests ECB mode (EcbInSymmCryptoCorrected), the code initializes the cipher in CBC mode (AES/CBC/PKCS5Padding)}. This inconsistency could lead to confusion and potential vulnerabilities''.


\begin{shaded*}
\noindent
\faInfoCircle{} \textit{We noted the absence of these test cases in CryptoAPI-Bench:\\
1. Test cases with unspecified encryption modes for Cipher, which default to using ECB, a mode known to be insecure.\\
2. Test cases with Cipher Block Chaining (CBC) encryption mode in client/server scenarios. This oversight may expose systems to security vulnerabilities, including the risk of an oracle-padding attack}
\end{shaded*}

We observed that ChatGPT successfully detected that the default encryption mode for symmetric encryptions, when not explicitly specified, is ECB and flagged it as insecure.

While the CBC mode offers a robust mechanism for data confidentiality through its probabilistic encryption properties, it lacks message integrity protection. This vulnerability renders CBC unsuitable for applications requiring both confidentiality and authenticity, particularly in client-server communication scenarios. An evaluation of ChatGPT's responses to code snippets that utilize CBC encryption mode, extracted from accepted answers on StackOverflow, reveals a tendency for the model to recommend CBC as a secure option. This occurs even when CBC may not be the appropriate choice. For instance, in a security assessment for a client/server code snippet (from SO post ID 18291987), ChatGPT continued to suggest using CBC, despite its known unsuitability in certain contexts.\footnote{https://chatgpt.com/share/0e74ad61-faf3-4b02-aa10-5551cc9ecf13}

\subsubsection{Predictable Key:}
While ChatGPT is not primarily designed as a detection tool, it notably surpassed CryptoGuard, a widely-used detection tool, in the realm of predictable key detection. ChatGPT achieved an impressive F-measure of 92.43\%, outperforming CryptoGuard, which attained a score of 76.92\%.

\subsubsection{Static Salt:}
\label{StaticSalt}
ChatGPT, with an F-measure of 88.6\%, demonstrates a slightly better ability to detect static salt than CryptoGuard, which has an F-measure of 85.71\%.

\begin{shaded*}
\noindent
\faInfoCircle{} \textit{We discovered the lack of test cases for small-size salts, a factor contributing to the weakness of Password-Based Encryption (PBE). Salts less than 64 bits are insufficient for effective security, as they significantly increase the risk of brute-force and rainbow table attacks.}
\end{shaded*}

We observed that ChatGPT can detect small-sized salts; for example, when directly provided with 4 as the salt size, it identifies this as a low salt size, which is insecure\footnote{https://chatgpt.com/share/0a0baca5-a2e1-4bd4-b621-b1acff4f2b06}. However, if the size is derived from a calculation, such as keysize/64 where keysize equals 256, it fails to detect the insecurity\footnote{https://chatgpt.com/share/a2163b80-7985-4412-9455-f98b4b64d066}.

\subsubsection{Low Iteration:}
While the comparison between ChatGPT and CryptoGuard may suggest a low detection ability for ChatGPT, in reality, we observed that ChatGPT demonstrates robust capability in detecting the value of iteration counts across various scenarios, even when passed through parameters. However, it is important to note that while ChatGPT can successfully identify the value of iteration count in many instances, there are cases where it struggles to detect insufficient numbers of iterations.

\subsubsection{Weak Random generation:}
ChatGPT exhibited slightly lower precision than CryptoGuard in detecting weak random generation but achieved better recall and F-measure.

\subsubsection{Predictable PBE Password:}
While ChatGPT exhibits proficiency in predictable key detection, its performance in predictable password detection is comparatively weaker.
\begin{shaded*}
\noindent\faInfoCircle{} \textit{We observed that there were no test cases for checking weak PBE algorithms, such as the insecure “PBEWithMD5AndDES” or the bad practice “PBKDF2WithHmacSHA1”. This absence of tests can lead to the usage of these vulnerable algorithms, thereby compromising the overall security of the encryption process..}    
\end{shaded*}

We observed that ChatGPT correctly flagged ``PBEWithMD5AndDES'' as a non-secure password-based key generation technique\footnote{https://chatgpt.com/share/cbc43c61-1dea-4131-afa2-6315e20d5f10}, but it recommended ``PBKDF2WithHmacSHA1'' as a secure option \footnote{https://chatgpt.com/share/9036dbd0-07dc-474a-8655-4c6e9c91437d}, whereas it is not a good practice. In some cases, ChatGPT suggested ``PBKDF2WithHmacSHA256'' for higher security.

\subsubsection{Static Initialization Vector:}

The precision of ChatGPT was comparable to CryptoGuard, but it demonstrated superior recall in this aspect (100\%).

\begin{shaded*}
 \noindent   \faInfoCircle{} \textit{We found no test cases to verify if the same IV (Initialization Vector) was reused. This oversight could lead to potential replay attacks or compromise the encryption scheme's security, making it easier for attackers to decrypt the data..}
\end{shaded*}

\subsubsection{Predictable Password in keyStore}
We observed that both ChatGPT and CryptoGuard demonstrated remarkable proficiency in detecting Predictable Passwords in keyStore. Their performance was notably similar, with both achieving a flawless 100\% recall and an F-measure exceeding 93\%.

\subsubsection{Vulnerability in SSL/TSL}
In our observations, ChatGPT did not perform as well as Cryptoguard. Cryptoguard outperformed it significantly, achieving a 100\% recall rate and an F-measure of 95.83\% in detecting SSL/TSL related vulnerabilities.

\subsubsection{Insecure Asymmetric Ciphers}
We observed that ChatGPT excelled in detecting RSA keys with a key size less than 2048 bits as insecure across various test cases, outperforming CryptoGuard in this aspect.

\begin{shaded*}
    \noindent\faInfoCircle{} \textit{We observed that there were no test cases for utilizing RSA with no padding, as well as for using PKCS1-v1.5 padding, both of which are considered bad practices in cryptography.}
\end{shaded*} 

We observed that ChatGPT can detect RSA with no padding\footnote{https://chatgpt.com/share/4e6bd975-7d05-440c-9c3e-0269006fc86a}. However, it incorrectly considers PKCS1-v1.5 padding as a secure and good option\footnote{https://chatgpt.com/share/9edfa3f1-c225-4b2e-b7c9-52a4f98d4fbe}, despite this being a bad practice and vulnerable to chosen-ciphertext attacks~\cite{ChosenCiphertext}.

\subsubsection{Weak Hash Function}
Similar to Cryptoguard, ChatGPT demonstrated proficiency in detecting Weak Hash Functions, achieving a recall rate of 100\%.

\begin{shaded*}
\noindent
\textit{In summary, we made the following observations about ChatGPT's performance in detecting cryptography misuse, highlighting areas where prompt engineering could be beneficial:
\begin{itemize}
\item High false positives in path-sensitive test cases.
\item Low accuracy in detecting insecure sizes.
\item Low recall and precision in detecting low iteration counts in PBE.
\item Considering CBC as a good practice regardless of context.
\item Considering common weak practices as good practices, such as using ``PBKDF2WithHmacSHA1'' for key generation or using ``PKCS1-v1.5'' padding.
\end{itemize}
}
\end{shaded*} 
    


 \begin{table*}[t]
 \caption{ChatGPT evaluation for basic and advanced test cases based on CryptoAPI-Bench. To ensure robust testing, we executed each test case three times. The total number of test cases, which is shown in parentheses, reflects this repetition.}
 \scalebox{0.94}{

\begin{tabular}{ll|cc|>{\centering\arraybackslash}p{1.2cm} >{\centering\arraybackslash}p{1.2cm} >{\centering\arraybackslash}p{1.2cm}|>{\centering\arraybackslash}p{1.2cm} >{\centering\arraybackslash}p{1.2cm} >{\centering\arraybackslash}p{1.2cm}|}
\cline{3-10}
\multicolumn{2}{l|}{\multirow{2}{*}{}}                                  & \multicolumn{2}{c|}{CryptoAPI-Bench}                     & \multicolumn{3}{c|}{ChatGPT}                                                 & \multicolumn{3}{c|}{ChatGPT with engineered prompts}                         \\ \cline{3-10} 
\multicolumn{2}{l|}{}                                                   & \multicolumn{1}{c|}{\#Positive Tests} & \#Negative Tests & \multicolumn{1}{c|}{TP}  & \multicolumn{1}{c|}{FP} & \multicolumn{1}{c|}{FN} & \multicolumn{1}{c|}{TP}  & \multicolumn{1}{c|}{FP}  & \multicolumn{1}{c|}{FN} \\ \hline
\multicolumn{2}{|l|}{Basic Cases}                                       & \multicolumn{1}{c|}{27 (81)}          & 13 (39)          & 78  & 9  & 3                       & 80  & 7  & 1                       \\ \hline
\multicolumn{1}{|l|}{\multirow{5}{*}{Advanced Cases}} & Interprocedural & \multicolumn{1}{c|}{40 (120)}         & 0                & 109 & 0  & 11                      & 118 & 0  & 2                       \\ \cline{2-10} 
\multicolumn{1}{|l|}{}                                & Field Sensitive & \multicolumn{1}{c|}{19 (57)}          & 0                & 44  & 0  & 13                      & 56  & 0  & 1                       \\ \cline{2-10} 
\multicolumn{1}{|l|}{}                                & Path Sensitive  & \multicolumn{1}{c|}{0}                & 20 (60)          & 0   & 54 & 0                       & 0   & 19 & 0                       \\ \cline{2-10} 
\multicolumn{1}{|l|}{}                                & Miscellaneous   & \multicolumn{1}{c|}{10 (30)}          & 2 (6)            & 24  & 0  & 6                       & 26  & 0  & 4                       \\ \cline{2-10} 
\multicolumn{1}{|l|}{}                                & Multiple Class  & \multicolumn{1}{c|}{20 (60)}          & 0                & 56  & 0  & 4                       & 59  & 0  & 1                       \\ \hline
\end{tabular}

}
\label{tab:BasicvsAdvanced}
 \end{table*}

\subsection{Prompt Engineering}

Despite the high volume of recent publications on large language models (LLMs), we found only four particularly useful for extracting prompt engineering techniques~\cite{white2023prompt, white2024chatgpt,schmidt2023cataloging,ekin2023prompt}.
From these papers, we learned that creating effective ChatGPT prompts necessitates understanding the user's objectives, recognizing ChatGPT's capabilities and limitations, incorporating relevant domain knowledge, and providing clear and specific instructions. Additionally, providing a few example responses can greatly assist ChatGPT in delivering more accurate and targeted answers.
\label{subsec:RQ2}

\subsubsection{Prompt Optimization:}
In order to gain more accurate detection results we did the following steps.

\emph{Role Specification:} We directed ChatGPT to assume the role of a security expert with extensive expertise in cryptography. This approach enhances the accuracy and relevance of the responses by focusing on a specific domain.

\emph{Clarification:} For cases where we find that ChatGPT recognizes common bad practices as good or practices that can be good or bad depending on the context (such as using CBC), we clearly state that these practices are blacklisted and issue a warning against them.

\emph{Guided Logical Processing:} We implemented steps to refine ChatGPT's understanding of checking the outcomes of calculations and conditions. This adjustment has improved the model's ability to detect false positives, enhance path sensitivity, and incorporate the results of calculations into its security assessments. For example, in section 4.1.4, we discussed how previously ChatGPT could not detect an insecure salt size when calculating 
keysize/64 where the keysize equals 256. However, with these improvements, it can now recognize such issues\footnote{https://chatgpt.com/share/179f3f49-6357-489f-994f-01bf193f7c30}.

\emph{Providing Code Examples:} 
To minimize ambiguity and enhance the precision of ChatGPT's analysis, few-shot examples are provided to clarify specific issues, thereby improving ChatGPT's ability to offer accurate and actionable advice.

\emph{Guided Breakdown of Rules:} To enhance ChatGPT’s proficiency in detecting misuses of the JCA, each violation rule is meticulously broken down into simpler, more manageable components. This structured approach allows ChatGPT to better understand and evaluate each element.

 \begin{lstlisting}[caption=Example of a true-negative test case, label=lst:TNTestCase]
public class BrokenCryptoABPSCase4 {
    public void go(int choice) throws NoSuchPaddingException, NoSuchAlgorithmException, InvalidKeyException {
        KeyGenerator keyGen = KeyGenerator.getInstance("AES");
        SecretKey key = keyGen.generateKey();
        Cipher cipher = Cipher.getInstance("RC2");
         if(choice>1){
            cipher = Cipher.getInstance("AES/CBC/PKCS5Padding");
            cipher.init(Cipher.ENCRYPT_MODE, key);
     }
    public static void main (String [] args) throws NoSuchPaddingException, NoSuchAlgorithmException, InvalidKeyException {
        BrokenCryptoABPSCase4 bc = new BrokenCryptoABPSCase4();
        int choice=2;
        bc.go(choice);
    }
}
\end{lstlisting}
\begin{lstlisting}[caption= Optimized prompt result for Listing 1, backgroundcolor=\color{gray!10},basicstyle=\color{black}\ttfamily\footnotesize ,label=lst:PromptResult, numbers=none]
Encryption Algorithms:
Number of algorithms=2,
Algorithms= {[AES, Suggested Algorithm], [RC2, Insecure Algorithm]} 
-------------------------
Encryption Modes:{ [CBC, CBC (insecure in client/server mode)], [NotSpecified, insecure ECB]}
--------------------------
Encryption Algorithm/Mode used in the main:[AES][CBC] , [Suggested Algorithm],[CBC (insecure in client/server mode)] 
\end{lstlisting}

\emph{Control of Verbosity:} A recent study \cite{Kabir} indicates that ChatGPT often provides verbose replies, which can result in the transmission of incorrect information. To address this issue, we have adjusted the verbosity of ChatGPT's responses. This adjustment aims to reduce ambiguity and the spread of misinformation.

Listing \ref{lst:PromptResult} shows ChatGPT's response \footnote{https://chatgpt.com/share/67a850f6-aa1d-4527-9bf8-1f2378ac8c1a} alongside the guided prompt for the code snippet in Listing \ref{lst:TNTestCase}, which represents a path-sensitive and true negative test case. Initially, ChatGPT flagged it as insecure, struggling to track the data flow effectively. However, with the assistance of guided prompts, ChatGPT's ability to detect path-sensitive test cases improves significantly. Additionally, CryptoGuard and CogniCrypt also flagged it as insecure \cite{AfroseDetectionTools}.

\subsubsection{Result:}
Table \ref{tbl:compare} shows the results of optimized prompts on CryptoAPI-Bench test cases and demonstrates that ChatGPT with optimized prompts has superior detection capabilities and outperforms CryptoGuard in security risk detection across all cryptographic categories.
Table \ref{tab:BasicvsAdvanced} 
details ChatGPT's performance, with and without prompt engineering, across both basic and advanced test cases.

To ensure that our optimizations extend beyond CryptoAPI-Bench, we also tested the effectiveness of our optimized prompts using CAMBench (34 new test cases). We specifically evaluated areas that were previously identified as weak and achieved an F-measure of over 90\%.

\section{Discussion}
\label{sec:Discussion}

\subsection{ChatGPT vs. Static Analysis Tools}

\emph{Parameter Passing}. Static analysis tools can struggle when it comes to accurately tracking parameter passing in code \cite{AutoDetecJavaCrypto, b3}. This can lead to missed vulnerabilities or errors. However, ChatGPT has demonstrated proficiency in understanding parameter passing (see Section \ref{StaticSalt} example). 

\emph{Lower and Upper Case Usage}.
Static analysis tools often face challenges in handling variations in case sensitivity \cite{b3}. ChatGPT, on the other hand, can adeptly detect vulnerabilities regardless of case usage. For example, whether it is ``AES/ECB'' or ``aes/ecb'', ChatGPT is capable of identifying vulnerabilities in both cases.

\emph{Path Sensitivity}.
Static tools may struggle with path-sensitive cases, where the behavior of the program depends on the specific path taken through the code \cite{AfroseDetectionTools,b3}. While initial challenges may exist with path-sensitive cases, ChatGPT can overcome them through prompt engineering and guidance (see Table  \ref{tab:BasicvsAdvanced}). 

\subsection{ChatGPT and Iteration Count}
While investigating the ability of ChatGPT to detect iteration counts less than 1000 for PBE in JCA, we observed that it struggled with identifying low iteration counts in several cases. We attempted to address this challenge by providing prompt guidance and using a few-shot example approach. However, despite detecting the iteration count correctly in some instances, ChatGPT still produced incorrect results in comparison. For instance, in the answer\footnote{https://chatgpt.com/share/70462df8-dfc6-4e3c-8d95-1854e84dc48f} for one of the test cases, it correctly identified an iteration count of 1020 but incorrectly reported it as less than 1000.

\section{Threats to Validity}
\label{sec:ThreatstoValidity}

The possibility of data leakage from CryptoAPI-Bench, which may have been used in ChatGPT's training, is a potential threat to internal validity.
However, there is no descriptive information about the misuses linked to the test cases (code snippets) and the issues are detailed in a separate Excel file located in a different directory. 
ChatGPT's descriptions did not closely resemble the content of the Excel file, and the low precision in detecting misuse for low Iteration count cases (48\% precision)  suggests that data leakage is unlikely. Nevertheless, further investigation is warranted.

The variability in ChatGPT's responses, even when the same prompt is used multiple times, also poses a potential threat to internal validity. To mitigate this threat, we implemented a procedure where each prompt was submitted three times. We then based our analysis on the aggregated results of these three responses, rather than relying on a single instance. This approach helps to ensure a more reliable assessment of ChatGPT's detection ability.

There exists a threat to external validity from the fact that we mainly relied on the CryptoAPI-Bench benchmark to assess ChatGPT, limiting the generalizability of our findings. To address this issue, we also analyzed a set of misuse cases from StackOverflow (in total 24 test cases) that were not addressed in CryptoAPI-Bench.
Additionally, we examined optimized prompts on test cases from CAMBench (34 new test cases).

\section{Conclusion}
\label{sec:Conclusion}

Cryptography is crucial for data confidentiality, but the complexity of cryptography APIs often leads to widespread misuse in software systems.  

We investigated ChatGPT’s performance in detecting cryptography misuses using two cryptography benchmarks and 24 test cases collected from StackOverflow. Our findings revealed that, with optimized prompts, ChatGPT is effective at identifying cryptography misuses in Java Cryptography Architecture (JCA). Notably, it outperforms CryptoGuard, the leading state-of-the-art cryptography misuse detector.

We conducted this study using GPT-3.5 Turbo, and in the future, we plan to investigate current models such as GPT-4o for ChatGPT and cover test cases that are representative of real-world cryptography API use.

\newpage

\bibliographystyle{ACM-Reference-Format}
\bibliography{ESEM2024}

\end{document}